\documentclass[aps,prb,showpacs,twocolumn,superscriptaddress,longbibliography]{revtex4-2}
\usepackage{amsfonts}
\usepackage{amssymb}
\usepackage{amsmath}
\usepackage{graphicx}
\usepackage{epsfig}
\usepackage{subfigure}
\usepackage{appendix}
\usepackage{hyperref}
\usepackage{url}

\hypersetup{hypertex=true,
colorlinks=true,
linkcolor=blue,
urlcolor = blue,
citecolor=blue}

\begin{document}

\title{Critical dynamics and superconducting state preparation in the quenched Kitaev chain with pairing imbalance}
\author{Y. B. Shi}
\email{1120210058@mail.nankai.edu.cn}
\affiliation{School of Physics, Nankai University, 300071 Tianjin, China}
\affiliation{School of Physics, Peking University, 100871 Beijing, China}
\author{Y. X. Zhang}
\affiliation{Arnold Sommerfeld Center for Theoretical Physics,
Ludwig-Maximilians-Universit\"at M\"unchen, Theresienstr. 37, 80333 M\"unchen, Germany}
\author{S. W. Liu}
\affiliation{School of Physics, Peking University, 100871 Beijing, China}
\author{Z. Song}
\email{songtc@nankai.edu.cn}
\affiliation{School of Physics, Nankai University, 300071 Tianjin, China}

\begin{abstract}
The dynamical balance of the pairing term plays a crucial role in the emergence of topological superconductivity in the $p$-wave spinless Kitaev chain, particularly in the non-Hermitian regime.
In this work, we systematically investigate the effects of non-Hermitian pairing terms on both equilibrium and nonequilibrium phenomena in the Kitaev chain. 
Our analysis focuses on two representative forms of pairing imbalance: uniform and staggered. We demonstrate that a uniform imbalance induces only minor perturbations to the spectrum and dynamical properties, without significantly affecting its equilibrium phase or nonequilibrium steady behavior. 
In contrast, even a slight staggered imbalance leads to drastic changes. 
At the symmetry point, it enables the resonant generation of two distinct superconducting states through critical dynamics, with the realized state determined by the direction of the bias.
Both states exhibit exact off-diagonal long-range order (ODLRO) in the thermodynamic limit. 
Our results emphasize the fragility of coherent dynamics in non-Hermitian topological systems 
and elucidate the interplay among non-Hermiticity, topology, and dynamical criticality in quench processes.
\end{abstract}

\maketitle

\section{Introduction}

An important axiom of quantum mechanics requires that for every physical observable, such as energy or momentum, there exists a linear and Hermitian operator that acts on the states of a closed system and produces the expectation value of the corresponding observable. 
The Hermiticity of the Hamiltonian guarantees the unitarity of the evolution operator, thereby ensuring probability conservation in isolated quantum systems~\cite{messiah2014quantum}. 
However, in realistic situations, physical systems are inevitably coupled to their environments, leading to effective probability nonconservation. 
Non-Hermitian Hamiltonians~\cite{Bergholtz2021Exceptionala,ashida2020non} have thus emerged as effective descriptions of open systems, establishing a new paradigm for interpreting and manipulating emergent phenomena absent in their Hermitian counterparts~\cite{hatano1996localization,Yao2018Edge,Rudner2009Rudner,Hu2011Absence,Borgnia2020Non,Lee2019Topological,Chen2020Emergent,yang2025quantum,bomantara2025nonhermitian,shimomura2025subspace}, such as the occurrence of exceptional points~\cite{bender2007making} and the non-Hermitian skin effect~\cite{Yao2018Edge}.
As expected, the dynamics governed by non-Hermitian Hamiltonians display a rich variety of
unique phenomena and have attracted significant attention in both
theoretical and experimental studies~\cite%
{Wang2019NonHermitian,Zhang2021Quantuma}.
Previous works have mainly focused on single-particle non-Hermitian Hamiltonians, such as nonreciprocal hopping~\cite{Fukui1998Breakdown,Shi2022Exceptional} or imaginary on-site potentials~\cite{Esaki2011Edge,Yokomizo2019Bloch}, whereas studies of systems without particle-number conservation remain scarce~\cite{Sayyad2023Topological}. Nevertheless, many-body non-Hermitian evolution can be experimentally realized, for example, by systematically post-selecting ancilla qubits~\cite{shen2025observation} or employing monitored quantum circuits~\cite{Fleckenstein2022non}.

The Kitaev chain~\cite{kitaev2001unpaired,Dvir2023Realization} is a well-studied
model where the parity of the particle number is conserved but particle number itself is not conserved. 
This model has been introduced to describe one-dimensional topological
superconductors with $p$-wave spinless pairing and acts as the fermionized version of
the famous one-dimensional transverse-field Ising model, which is one of
the simplest solvable models exhibiting quantum criticality and
demonstrating a quantum phase transition with spontaneous symmetry breaking~\cite{Pfeuty1970onedimensional}. 
The topological superconducting phase can be predicted by unpaired Majorana
modes localized at the ends of an open Kitaev chain. 
Recent experimental demonstrations~\cite{Dvir2023Realization,ten2024two} of bottom-up engineering of a Kitaev chain have confirmed that a minimal Kitaev chain can be realized using only two quantum dots coupled via a superconductor~\cite{Leijnse2012Parity}, and that Majorana edge modes emerge at finely tuned sweet spots without topological protection~\cite{cayao2025non}.
Pair creation and annihilation terms play a crucial role in the existence of the superconducting phase, where spinless fermions with opposite momenta can pair up and form superconducting Cooper pairs.~\cite{vodola2014kitaev,vodola2015long,viyuela2016topological,lepori2017long,bhattacharya2019critical,koffel2012entanglement,hauke2013spread,grass2014trapped,dutta2017probing,vajna2015topological,sedlmayr2018bulk,maslowski2023dynamical,cheraghi2023dynamical,sim2022quench,molignini2017sensing,molignini2018universal,molignini2020edge,Zeinab2025Electronic}. 
Recent studies show that imbalanced pairing terms preserve both the superconducting phases and their topological features~\cite{Li2018Topologicala}, with the bulk–boundary correspondence remaining robust against non-Hermitian effects.
Nevertheless, there are complex Bogoliubov-de Gennes (BdG) spectra associated with $\mathcal{PT}$-symmetry breaking in Kitaev chains with imaginary $p$-wave pairing \cite{yang2020resonant} predicting the existence of the exceptional points.
Moreover, non-Hermitian Kitaev chains with staggered imbalanced pair
creation and annihilation terms \cite{Shi2023Fixed,Shi2023Topological}
exhibit far richer phase diagrams, which further complicate the investigation into the dynamical
behaviors of such systems.

In this paper, we investigate the influence on the quench dynamics~\cite{heyl2013dynamical,heyl2015scaling,Unal2020Topological,Entanglement2024Vimal} induced by two types of non-Hermitian pairing impurities in a Kitaev chain. 
A uniform pairing imbalance neither shifts the equilibrium phase boundary nor induces imaginary eigenenergies. 
Although the norm of the evolved state is not conserved during the quench dynamics, the order parameter, which is originally introduced to characterize the pairing rate in the Hermitian Kitaev chain~\cite{Shi2022Dynamic}, becomes stable after a relaxation time. 
The steady behavior of order parameter can also be used to map out the quantum phase diagram of the post-quench Hamiltonian, in a manner analogous to Landau-type dynamical phase transitions~\cite{Homrighausen2017Anomalous} in Hermitian systems.
However, a staggered pairing imbalance can not only shift the topological phase boundary but also break the time-reversal $\left( \mathcal{T}\right) $ symmetry when the chemical potential is small. 
The nonequilibrium steady order parameter remains valid in this case and increases rapidly as $\lvert \mu \rvert$ decreases within the $\mathcal{T}$-symmetry-broken (TSB) phase.
We find that two distinct superconducting states can be generated via the critical dynamics~\cite{yang2020resonant} and the realized state is determined by the direction of the bias at the symmetry point.
Remarkably, these states exhibit exact off-diagonal long-range order (ODLRO)~\cite{Yang1989eta,yang1990so}, which is a hallmark of superconductivity and can be rigorously established only from the exact many-fermion wavefunction in the thermodynamic limit.
Our findings thus offer an alternative route to generate a superconducting state far from the ground state through a dynamical process, as opposed to lowering the temperature or employing imaginary-time evolution.

The rest of the paper is organized as follows. Section~\ref{Model and quench dynamics} introduces the Kitaev model with non-Hermitian pairing terms and the quench protocol employed in this work. In particular, Section~\ref{Hermitian Kitaev chain and pairing in $k$ space} presents the phase diagram and dynamical properties of the Hermitian Kitaev chain. Sections~\ref{Uniform pairing imbalance} and~\ref{staggered pairing imbalance} investigate the effects of uniform and staggered pairing imbalances, respectively. In Section~\ref{Dynamical generation of superconducting state with ODLRO}, we study the states generated by the staggered imbalance at the symmetry point and demonstrate the existence of ODLRO associated with the condensation of zero-momentum fermion pairs. Finally, Section~\ref{Summary} summarizes our results and provides a discussion.

\begin{figure}[t!]
\centering\includegraphics[width=0.48\textwidth]{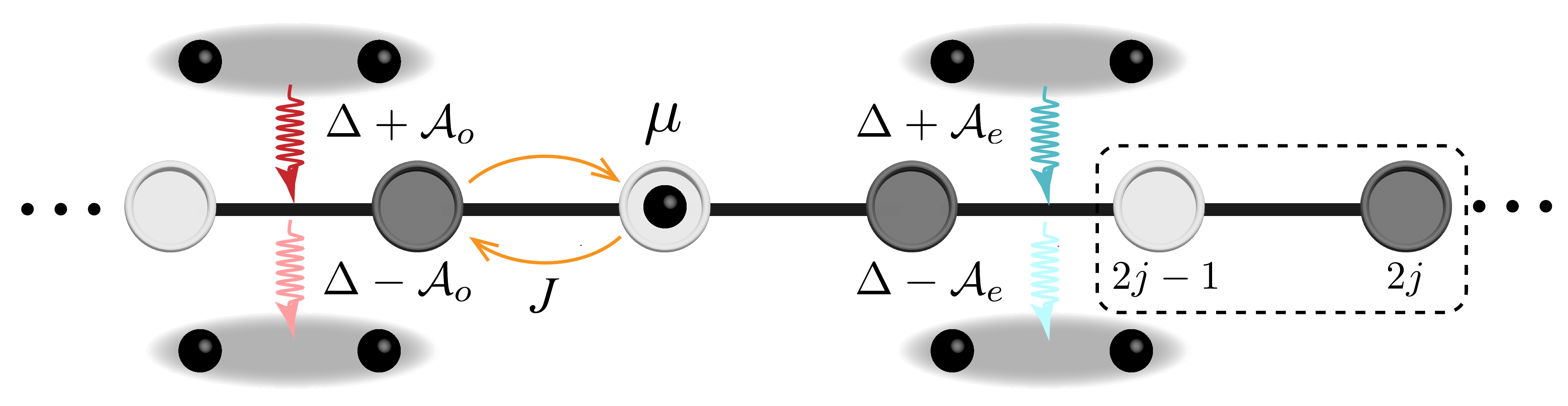}
\caption{
Schematic illustration of the 1D Kitaev model for spinless fermions with a non-Hermitian pairing imbalance.
Here, $J$ and $\Delta$ denote the hopping amplitude and pairing potential, respectively, and $\mu$ is the chemical potential. $\mathcal{A}_o$ ($\mathcal{A}_e$) represents the imbalanced pairing amplitude on the odd (even) dimer. The unit cell consists of two sublattices, shown in white and gray, as indicated by the dashed black rectangle.}
\label{fig1}
\end{figure}

\begin{figure*}[t!]
\centering\includegraphics[width=0.9\textwidth]{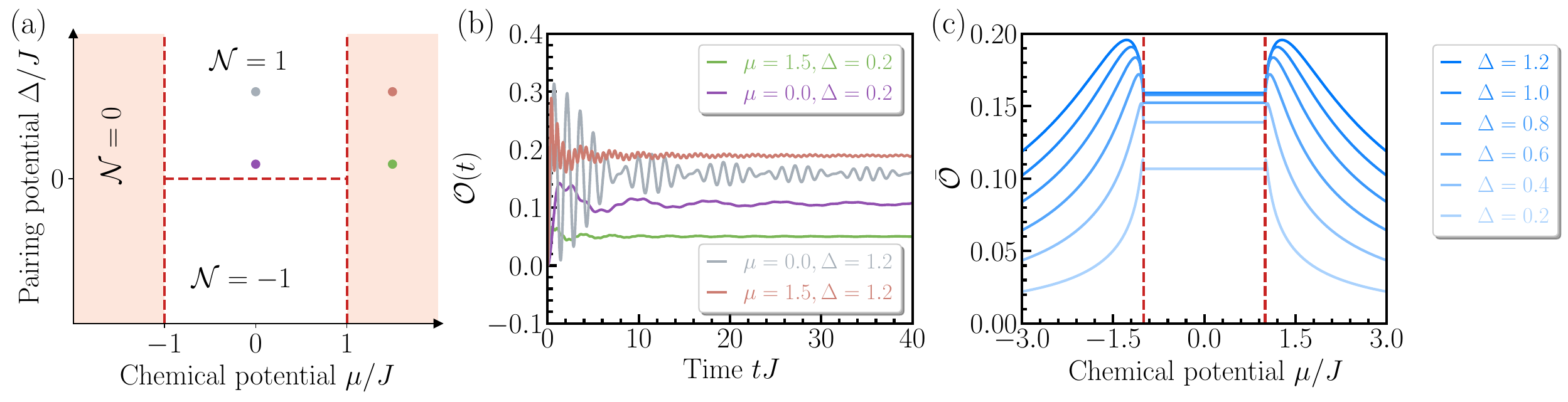}
\caption{
(a) Phase diagram of the Hermitian Kitaev chain [cf. Eq.~\eqref{eq:Kitaev_chain}] on the $\mu$–$\Delta$ plane. The pink (white) regions correspond to the trivial (non-trivial) topological phase. Red dashed lines indicate the phase transition boundaries. 
The four colored dots indicate representative parameter sets, labeled in the legend of (b) with the corresponding colors.
(b) Profiles of instantaneous order parameters as a function of time with four sets of parameters ($\mu$ and $\Delta$) indicated in (a). They all exhibit damping oscillation around their stable values. 
(c) Profiles of the time-averaged order parameter $\overline{\mathcal{O}}$, defined in Eq.~\eqref{eq:time_averaged_order_parameter}, along lines of varying $\Delta$. They indicate that $\overline{\mathcal{O}}$ is non-analytic at the phase boundary, which is labeled by the red dashed lines. Other parameters are $N = 4000$ and $J = 1$.
}
\label{fig2}
\end{figure*}

\section{Model and quench dynamics}
\label{Model and quench dynamics}

In this section we introduce the general form of the non-Hermitian dimerized
Kitaev chain $H$ on a lattice of length $2N$. In general, a non-Hermitian Hamiltonian can be decomposed into Hermitian and anti-Hermitian components,
\begin{equation}
H=H_{\mathrm{0}}+H_{\mathrm{NH}},  \label{Eq:model}
\end{equation}
where
\begin{eqnarray}
    H_{0}= \frac{ H+H^{\dagger }}{2}\textrm{, and } H_{\mathrm{NH}}= \frac{ H-H^{\dagger }}{2}.
\end{eqnarray}
Concretely, $H_{0}$, which represents a Hermitian Kitaev chain, takes the form%
\begin{equation}
H_{\mathrm{0}}=\sum_{j=1}^{2N}Jc_{j}^{\dagger }c_{j+1}+\Delta c_{j}^{\dagger
}c_{j+1}^{\dagger }+\mathrm{H.c.}+\mu \left( 1-2n_{j}\right). \label{eq:Kitaev_chain}
\end{equation}%
Here, $J$ denotes the hopping amplitude, $\Delta$ the pairing potential, and $\mu$ the chemical potential. The operator $c_j^\dagger$ creates a spinless fermion at site $j$, so each site can be either empty or singly occupied, naturally forbidding double occupancy. The corresponding fermion density operator is defined as $n_j = c_j^\dagger c_j$. We impose periodic boundary conditions (PBC) by setting $c_{2N+1} = c_1$, and open boundary conditions (OBC) by setting $c_{2N+1} = 0$.

The second $H_{\mathrm{NH}}$, which introduce the imbalanced dimerized pairing imbalance, can be expressed as
\begin{equation}
H_{\mathrm{NH}}=\sum\limits_{j=1}^{N}\mathcal{A}_{\mathrm{o}}c_{2j-1}^{\dagger
}c_{2j}^{\dagger }+\mathcal{A}_{\mathrm{e}}c_{2j}^{\dagger }c_{2j+1}^{\dagger }-%
\mathrm{H.c.},
\end{equation}%
Here, $\mathcal{A}{\mathrm{o}}$ ($\mathcal{A}{\mathrm{e}}$) represents the pairing imbalance on the odd (even) dimers. In this work, the amplitudes considered are much smaller than both the hopping strength and the pairing potential, mathematically expressed as
\begin{eqnarray}
    \mathcal{A}_{\mathrm{o}},\mathcal{A}_{\mathrm{e}}<J\text{, and, } \Delta.
    \label{eq:condition}
\end{eqnarray}
All parameters considered in this work are real. A schematic illustration of the model is presented in Fig.~\ref{fig1}.
Before calculating the spectrum and investigating the dynamical properties, it is profitable to investigate the symmetry of the system and its
spontaneous breaking in the eigenstates first.
The first symmetry is fermion-number parity, expressed as $\left[ \Pi ,H\right] =0$, where the parity operator is defined as $\Pi=\prod_{j=1}^{2N}\left( -1\right) ^{n_{j}}$. 
The second symmetry is time-reversal $\left( \mathcal{T}\right) $, where the anti-linear time-reversal operator $\mathcal{T}$ satisfies $\mathcal{T}i\mathcal{T}=-i$. 
In non-Hermitian models, time-reversal symmetry plays a
role analogous to $\mathcal{PT}$ symmetry, as both guarantee that
the eigenvalues are either real or occur in complex-conjugate pairs. 
The third symmetry is translational invariance, which allows the Hamiltonian to be block-diagonalized.

In this paper, we consider the following quench protocol: We prepare the initial state as the empty state $\vert \psi(0)\rangle = \vert 0\rangle$. 
In a quantum quench experiment, such a state can be prepared in the ground state for infinite potential, i.e., $\mu\rightarrow \infty$. 
Next, we introduce the post-quench Hamiltonian $H$ and for any given time $t$, the evolved state takes the form
\begin{eqnarray}
\label{eq:evloved_state}
    \left \vert
\psi \left( t\right) \right \rangle = \exp 
\left( -iHt \right)
\vert 0 \rangle. 
\end{eqnarray}
Since the Hamiltonian $H$ can be exactly block diagonalized via the Fourier transformation, the evolved state is always expressible as a matrix product state in momentum space, $  \left\vert
\psi \left( t\right) \right\rangle = \prod_k \vert \psi(t)\rangle_k$, during the evolution. 

Dynamical phase transitions~\cite{heyl2013dynamical,Shi2022Dynamic,Homrighausen2017Anomalous} have emerged over the past decade as nonequilibrium analogs of equilibrium phase transitions.
One perspective resembles the Landau paradigm of equilibrium phase transitions, in which non-analytic or scaling behavior is sought in the dynamics of order parameters, two-point correlation functions, or response functions~\cite{Hashizume2022Dynamical}.
Specifically, one can measure a dynamical order parameter $\mathcal{O}(t)$ of an evolved state under quench dynamics. After a relaxation time, it approaches a steady value defined as
\begin{eqnarray}
    \mathcal{\bar{O}} =\lim_{T\rightarrow\infty} \frac{1}{T}  \int^{T}_{0} \mathcal{O}(t) \mathrm{d} t. 
\label{eq:time_averaged_order_parameter}
\end{eqnarray}
This nonequilibrium steady value is generally expected to capture the characteristics of the post-quench Hamiltonian or to reveal novel emergent properties~\cite{Shi2024Emerging}.

\section{Hermitian Kitaev chain and pairing in $k$ space}
\label{Hermitian Kitaev chain and pairing in $k$ space}
Prior to investigating the effects of non-Hermitian impurities, we first examine the properties and dynamical behavior of the Hermitian system. The model Hamiltonian $H$ represents the bare Kitaev chain with nearest-neighbor hopping and pairing when the non-Hermitian term $H_{\mathrm{NH}}=0$.
We can block diagonalize the Hamiltonian~\eqref{eq:Kitaev_chain} based on Fourier transformation,
$
    c_{j}={1}/{\sqrt{N}} \sum_{k}e^ {ikj} c_{k},
$
with wave vector $k\in (-\pi ,\pi ]$ in the large $N$ limit. 
Since both electrons and holes are involved, it is convenient to rewrite the momentum-space Hamiltonian in the Nambu basis,
\begin{equation}
H=H_k = \sum_{\pi >k>0}\psi _{k}^{\dagger }H_{\mathrm{BdG}}\psi _{k},
\end{equation}%
where $\psi _{k}=\left( c_{k},c_{-k}^{\dagger }\right) ^{T}$ defines a Nambu
operator and $H_{\mathrm{BdG}}$ is the {BdG} Hamiltonian given by
\begin{equation}
H_{\mathrm{BdG}}=2\left( 
\begin{array}{cc}
J \cos k-\mu  & i\Delta \sin k \\ 
-i\Delta \sin k & \mu - J \cos k%
\end{array}%
\right) .
\end{equation}%
The energy spectrum of $H_{\mathrm{BdG}}$ takes the form%
\begin{equation}
E_{k,\pm }=\pm 2\sqrt{\left( J\cos k-\mu \right) ^{2}+\Delta ^{2}\sin ^{2}k},
\label{Eq:energy_spectrum}
\end{equation}
and the energy gap closes at 
\begin{equation}
    \left\vert \mu \right\vert=J  \text{ and }\Delta=0\text{, for }\mu<J ,
\end{equation} 
signaling a topological phase transition.
The topological phase is characterized by the winding number $\mathcal{N}$~\cite{chiu2016classification}, which is illustrated in Fig.\ref{fig2}(a). In the topological phase, Majorana edge states emerge at the ends of the chain under OBC. These states are exponentially localized at the edges and decay into the bulk.

To gain an intuitive understanding of the behavior of the dynamical order parameter, we consider the Bardeen-Cooper-Schrieffer (BCS)-like pairing rate in momentum space. The operator is defined as
\begin{equation}
    \widehat{\mathcal{O }}_k= ic_{k}^{\dagger }c_{-k}^{\dagger }-ic_{-k}c_{k}. \label{eq:orderparameter}
\end{equation}
For a given state $\vert \phi \rangle$, the quantity $\vert \langle \phi \vert \hat{\mathcal{O}}_k \vert\phi\rangle \vert $ characterizes the pairing rate in the $k$ channel.
The average pairing rate of the time-evolved state is given by
\begin{eqnarray}
    \mathcal{O}(t) = \frac{1}{N}\sum_{\pi > k > 0} \vert \langle  \psi(t) \vert \hat{\mathcal{O}}_k \vert \psi(t)\rangle \vert.
\end{eqnarray}
As expected, $\mathcal{O}(t)$ becomes steady after a sufficiently long time, cf. Fig.~\ref{fig2}(b). 
The steady value of the dynamical order parameter can be obtained from
\begin{eqnarray}
    \overline{\mathcal{O}} = \frac{1}{N} \sum_{\pi >k >0 } \left \vert  \frac{(\cos k -\mu) \Delta \sin k }{(\cos k -\mu)^2 + \Delta^2 \sin^2 k } \right \vert. \label{steady_OP_hermitian}
\end{eqnarray}
In the thermodynamic limit, Eq.~\eqref{steady_OP_hermitian} can be expressed as the integral equations and the analytical expressions is given by
\begin{eqnarray}
    \overline{\mathcal{O}} = \left\{ \begin{array}{cc}
\zeta & \textrm{for, }  | \mu | \geq 1\\
\frac{|\Delta |\ln( |\Delta |)}{\pi \Delta ^{2} -1} & \textrm{for, }  | \mu | < 1
\end{array} \right. .
\end{eqnarray}
with 
\begin{eqnarray}
    \zeta &=& \frac{\vert \Delta \vert}{2\pi(1-\Delta^2)} \ln \left\vert \frac{\vert \mu \vert+1 }{\vert \mu \vert-1} \right \vert  \\
    &-& \frac{\Delta^2 \vert \mu \vert}{2\pi (1-\Delta^2)\sqrt{\mu^2+\Delta^2-1}} \ln \left\vert \frac{\sqrt{\mu^2+\Delta^2-1}+ \vert \Delta \vert }{\sqrt{\mu^2+\Delta^2-1}- \vert \Delta \vert} \right \vert. \notag 
\end{eqnarray}
Fig.~\ref{fig2}(c) depicts the profile of the steady order parameter $\overline{\mathcal{O}}$ as a function of $\mu$ for various values of $\Delta$.
The numerical results clearly demonstrate that $\overline{\mathcal{O}}$ effectively outlines the phase diagram, as non-analytic behavior emerges precisely at the phase boundaries indicated by the red dashed lines. Notably, in the topological phase, the order parameter exhibits a plateau whose value depends solely on $\Delta$. 
Hence, a natural question arises: do these results remain valid, or are they obscured when the system becomes non-Hermitian?

\begin{figure}[t!]
\centering\includegraphics[width=0.48\textwidth]{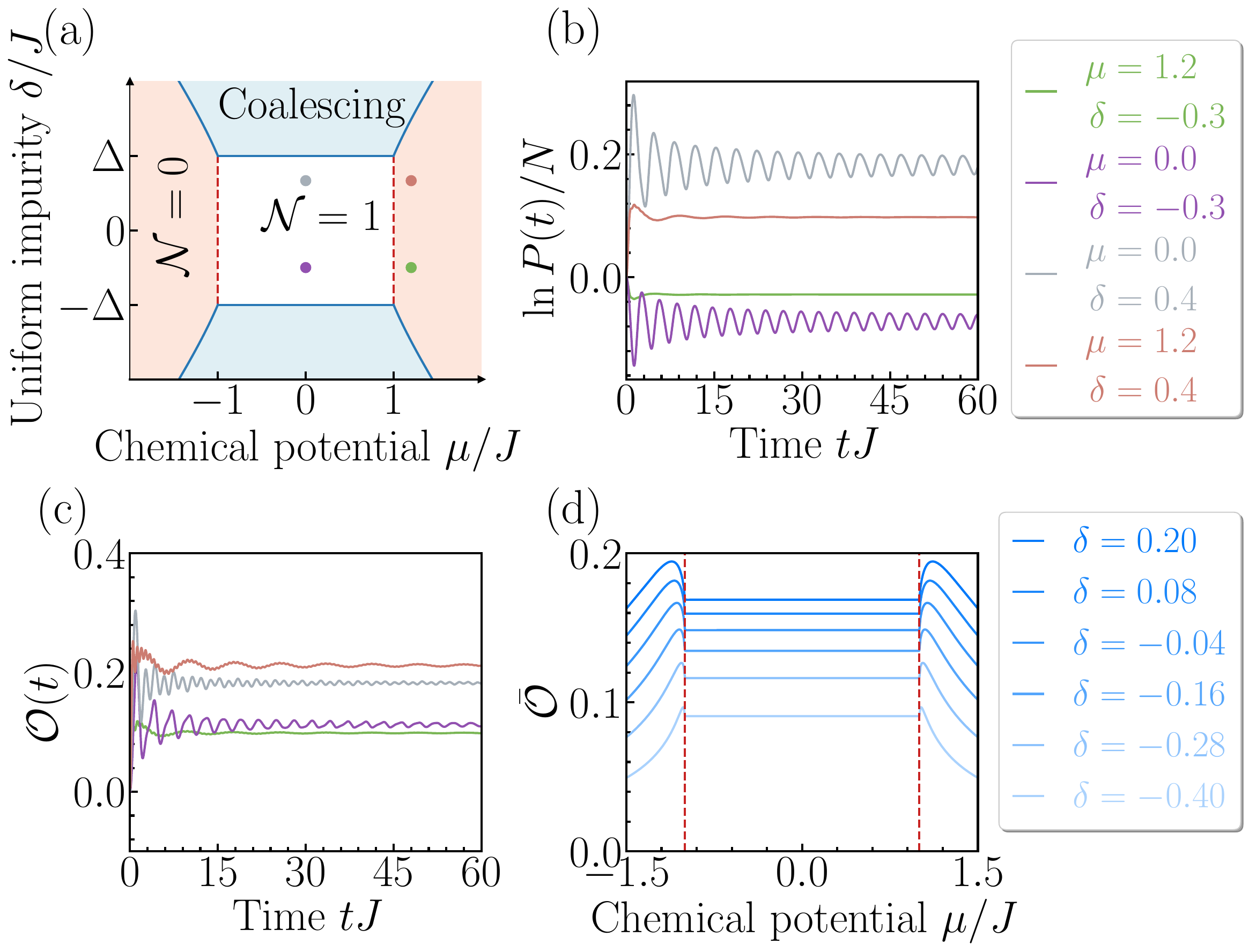}
\caption{
(a) Phase diagram of non-Hermitian Kitaev chain with uniform pairing imbalance. The system is in trivial phase when $\vert \mu \vert > 1$ and $\mu^2 + \Delta^2 - J^2 > \delta^2$ (pink); in topological phase when $\vert \mu \vert < 1$ and $\Delta^2 > \delta^2$ (white); in coalescing phase with the presence of EP otherwise (cyan).
Blue solid curves and red dashed lines denote the $\mathcal{T}$-symmetry-breaking boundary and the topological phase boundary, respectively.
Profiles of (b) $\ln P(t)/N$ and (c) $\mathcal{O}(t)$ as a function of time $t$ with several typical sets of parameters ($\mu$ and $\delta$). 
(d) Plots of the steady order parameter $\overline{\mathcal{O}}$ as a function of the chemical potential $\mu$ for several small values of $\delta$. The data are obtained from Eq.~\eqref{eq:averaged_order_parameter}. The curves exhibit non-analytic behavior at the topological phase boundary, consistent with the Hermitian case. Other parameters are $J=1$, $\Delta=0.6$, and $N=4000$. 
}
\label{fig3}
\end{figure}
\section{Uniform pairing imbalance}
\label{Uniform pairing imbalance}
Having taken the Hermitian case as a baseline and benchmarked the pairing rate as an order parameter for the phase transition in nonequilibrium dynamics, we now turn to the uniform imbalance case, i.e., $\mathcal{A}_{\mathrm{o}} = \mathcal{A}_{\mathrm{e}} = \delta$.
In this scenario, the system preserves single-site translational symmetry, satisfying $[T, H] = 0$, where the translation operator $T$ acts on the fermionic operators as $T c_j T^{-1} = c_{j+1}$. Mathematically, we introduce the following similarity transformation:
\begin{equation}
d_{j}=\left( \frac{\Delta +\delta }{\Delta -\delta }\right) ^{1/4}c_{j},\bar{%
d}_{j}=\left( \frac{\Delta -\delta }{\Delta +\delta }\right)
^{1/4}c_{j}^{\dagger },  \label{Eq:mapping}
\end{equation}%
which allows the Hamiltonian to be rewritten in the form
\begin{eqnarray}
H &=&\sum_{j=1}^{2N}J\overline{d}_{j}d_{j+1}+\mathrm{H.c.}+\mu \left(
1-2n_{j}\right)   \notag \\
&&+\sqrt{\Delta ^{2}-\delta ^{2}}\left( \bar{d}_{j}\bar{d}%
_{j+1}+d_{j+1}d_{j}\right) .
\end{eqnarray}%
Although $\bar{d}_{j}$ is not the Hermitian conjugate of the operator ${d}_{j}$, i.e., $\bar{d}_{j} \neq d_{j}^{\dagger}$, they satisfy the following anti-commutation relations
\begin{equation}
\left\{ d_{j},\bar{d}_{l}\right\} =\delta _{jl},\left\{ \bar{d}_{j},\bar{d}%
_{l}\right\} =\left\{ d_{j},d_{l}\right\} =0.
\end{equation}%
The BdG formalism in real space for this Hamiltonian, constructed with the biorthonormal Nambu operators $\bar {D}=\left( d_{j},\bar{d}_{j}\right)$ and $D=\left( 
\bar{d}_{j},d_{j}\right) $ for $j=1,2,\dots,2N$, takes the same form as that of the Hermitian Kitaev chain, after replacing the real pairing amplitude $\Delta $ by $\sqrt{\Delta
^{2}-\delta ^{2}}$. 
Consequently, the energy spectrum is directly obtained from Eq.\eqref{Eq:energy_spectrum} as
\begin{equation}
E_{k,\pm }=\pm 2\sqrt{\left( J\cos k-\mu \right) ^{2}+\left( \Delta
^{2}-\delta ^{2}\right) \sin ^{2}k}.
\end{equation}%
When $\Delta> \delta$, the gap closes at $\vert\mu \vert=1$, consistent with the Hermitian case. The system is topological when $\vert\mu \vert < 1$; otherwise, it is in the trivial phase.
On the other hand, when $\delta$ satisfies the following conditions,
\begin{eqnarray}
    \delta ^{2}>\mu ^{2}+\Delta ^{2}-J^2,\text{ and }\delta>\Delta,
\end{eqnarray}
imaginary energy levels emerge, leading to the breaking of $\mathcal{T}$ symmetry in the corresponding eigenstates.
Exceptional points (EPs) always exists insides these regions~\cite{Li2018Topologicala}. This means that there is always a critical momentum $k_c$ satisfying $E_{k_c,+} = E_{k_c,-}$, where the spectrum is real on one side and complex on the other.
The EPs, also called non-Hermitian degeneracies or branch points in the complex energy plane, occur where two complex eigenvalues and their corresponding eigenvectors coalesce, forming what is known as a coalescing level. These points represent special singularities unique to non-Hermitian systems and have wide applications in state preparation. Mathematically, the occurrence of an EP is associated with the emergence of a Jordan block. The two eigenvectors of a $2 \times 2$ Jordan block coincide and are self-orthogonal. 
In such a case, the usual notion of a well-defined ground state breaks down, as does the definition of the topological invariant. Therefore, we refer to these regions as the coalescing phase. The detailed phase diagram is shown in Fig.~\ref{fig3}(a). Overall, the spectrum remains real under the condition given in Eq.\eqref{eq:condition}.

The evolved states before normalization can be calculated analytically according to Eq.~\eqref{eq:evloved_state}
\begin{eqnarray}
\left\vert \psi (t)\right\rangle   &=&\prod_{\pi>k>0} \frac{\sin \left( E_{k,+}t\right)}{E_{k,+}}  \\
 &\times& \{2(\Delta +\delta)
 \sin k\left\vert 1\right\rangle _{k}\left\vert
1\right\rangle _{-k}  \notag \notag \\
&+&\left[ E_{k,+}\cot \left( E_{k,+}t\right) -2i(\mu -J \cos k)\right]
\left\vert 0\right\rangle _{k}\left\vert 0\right\rangle _{-k} \}.\notag
\end{eqnarray}
The probability is naturally not conserved over time due to the presence of the non-Hermitian term,
\begin{eqnarray}
    P(t) &=& \left\langle \psi (t) \vert \psi (t)\right\rangle   \\
   &=& \prod_{\pi>k>0} \bigg[\cos^2(E_{k+}t )  \notag \\
   &+& \frac{(\Delta+\delta)^2\sin^2k+(\mu-J \cos k)^2}{(\Delta^2-\delta^2)\sin^2k+(\mu-J \cos k)^2}\sin^2(E_{k+}t )\bigg].\notag
\end{eqnarray}
We simulate the density of the logarithmic probability for several representative system parameters, as shown in Fig.~\ref{fig3}(b). After a relaxation time, the probability exhibits damped oscillations around a steady value. Qualitatively, while the non-Hermitian term induces an imbalance between pair creation and annihilation, these processes undergo coherent cancellation. 
As a consequence, although the spectrum is slightly perturbed, the gap-closing point remains unchanged, and the topological phase boundaries are preserved.
In quench dynamics, the preserved $\mathcal{T}$ symmetry of the post-quench Hamiltonian ensures that the evolved state exhibits coherent dynamics rather than collapsing into a single eigenstate.

The stability observed during long-time evolution motivates us to investigate the behavior of the pairing rate $\mathcal{O}(t)$. The corresponding operator is defined consistently with Eq.~\eqref{eq:orderparameter}. Straightforward algebra then yields
\begin{eqnarray}
    \mathcal{O}(t) &= & \frac{1}{N} \sum_{\pi > k > 0} \left \vert \frac{\langle  \psi(t) \vert \hat{\mathcal{O}}_k \vert \psi(t)\rangle}{\langle  \psi(t) \vert \psi(t)\rangle} \right \vert  \\
     &=& \frac{1}{N} \sum_{\pi > k > 0}
     \frac{8(\Delta+\delta)\sin k(\mu-J\cos k)\sin^2(E_{k,+}t)}{E_{k,+}^2+8(\delta\Delta+\delta^2)\sin ^2k \sin^2( E_{k,+}t)}. \notag
\end{eqnarray}
Fig.~\ref{fig3}(c) present $\mathcal{O}(t)$ as a function
of time using the same parameters as in Fig.~\ref{fig3}(b). As expected, the quantity becomes steady
after a long period as well. 
Although deriving an analytical form at finite time is challenging, the steady order parameter becomes expressible as a summation in the infinite-time limit,
\begin{eqnarray}
\label{eq:averaged_order_parameter}
    \overline{\mathcal{O}} &=& \frac{1}{N}\sum_{\pi>k>0}\left\vert \frac{\mu-J \cos k}{\Delta \sin k}     \right \vert \\
     &&\times \left(1-\sqrt{\frac{(J \cos k- \mu)^2+(\Delta^2-\delta^2)\sin^2 k}{(J \cos k- \mu)^2+(\Delta+\delta)^2\sin^2 k}}\right).  \notag 
\end{eqnarray}
Numerical results are shown in Fig.~\ref{fig3}(d). We find that it continues to clearly identify the phase boundary, with a plateau persisting within the nontrivial phase, as observed in the Hermitian case. 
Thus, we conclude that uniform impurities do not induce significant changes on both equilibrium and nonequilibrium phenomena compared to the Hermitian case.

\begin{figure}[t!]
\centering\includegraphics[width=0.45\textwidth]{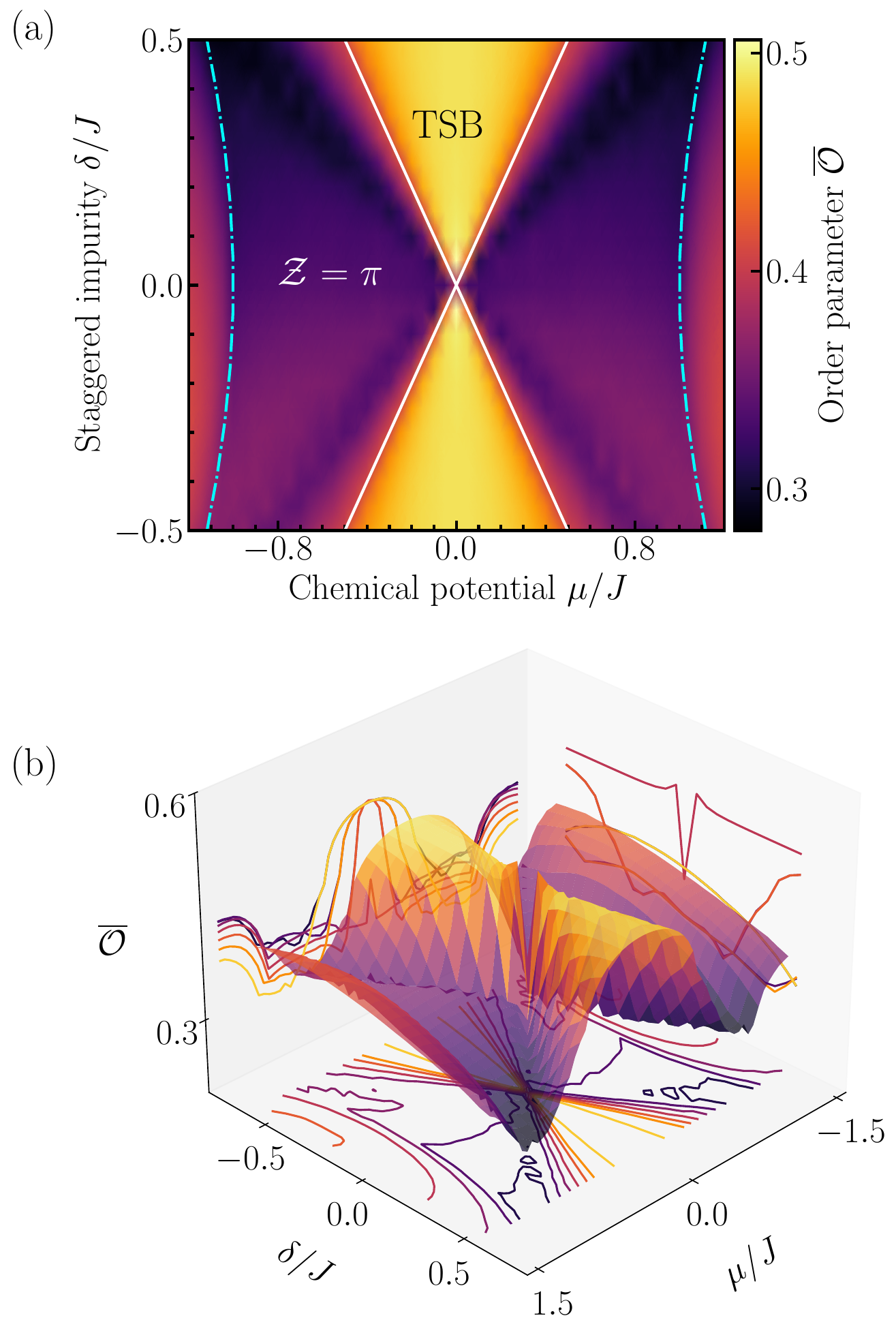}
\caption{(a) Two–dimensional contour color plot and (b) the corresponding three–dimensional surface with the contour projections of $\overline{\mathcal{O}}$ obtained from Eq.~\eqref{eq:order_parameter_staggered} as functions of $\mu$ and $\delta$. The white solid lines denote the $\mathcal{T}$-symmetry-breaking boundary, while the cyan dashed lines indicate the topological phase boundary. The order parameter exhibits non-analytic behavior along both boundaries. As $\mu$ decreases, the steady order parameter rapidly increases to a stable value. Other parameters are set as $J = 1$, $\Delta = 2$, and $N = 100$. The time step is set to $\delta t = 0.1$, and $1000$ steps are used for the quench dynamics.}
\label{fig4}
\end{figure}

\section{staggered pairing imbalance}
\label{staggered pairing imbalance}
The balance between pair creation and annihilation is completely destroyed when we consider a spatially inhomogeneous perturbation. To prove this, we introduce the staggered imbalance, where the intra- and inter-dimer pairing strengths are opposite, i.e., $\mathcal{A}_{\mathrm{o}}=-\mathcal{A}_{\mathrm{e}}=\delta$. 
In this case, the system exhibits a two-site translational symmetry,  $\left[ T,H\right] =0$, with the translation operator acting as $Tc_{j}T^{-1}=c_{j+2}$. Each unit cell thus consists of two sublattices, corresponding to the odd- and even-numbered sites.
One can block diagonalize the Hamiltonian by introducing the Fourier transformation on the two sublattices,
\begin{equation}
	\left( 
	\begin{array}{c}
		\alpha _{k} \\ 
		\beta _{k}%
	\end{array}%
	\right) =\frac{1}{\sqrt{N}}\sum_{k}e^{-ikj}\left( 
	\begin{array}{c}
		c_{2j-1} \\ 
		c_{2j}%
	\end{array}%
	\right),
\end{equation}
where $j=1,2,...,N$, $k=2m\pi /N$, $m=0,1,2,...,N-1$.
The BdG Hamiltonian of the momentum space Hamiltonian $H_k = \psi _{k}^{\dagger}H_{\mathrm{BdG}}\psi _{k}$ is given by 
\begin{equation}
	H_{\mathrm{BdG}}=\left( 
	\begin{array}{cccc}
		-2\mu & \gamma _{-k} & 0 &  \theta_k  \\ 
		\gamma _{k} & -2\mu &  -\theta_k^{\ast } & 0 \\ 
		0 & -\eta _{k}^{\ast } & 2\mu & -\gamma _{-k} \\ 
		\eta _{k} & 0 & -\gamma _{k} & 2\mu%
	\end{array}%
	\right) , \label{eq:BdG_Hamiltonian}
\end{equation}
with
\begin{eqnarray}
	\gamma _{k}&=&J+Je^{ik}, 
	\eta _{k} =\left( \Delta - \delta\right) -\left( \Delta + \delta\right) e^{ik}, \notag \\
    \theta_k &=& \left( \Delta + \delta\right)  -\left( \Delta - \delta\right)e^{-ik} .
\end{eqnarray}
The Nambu operator is $\psi _{k}^{\dagger }=\left(\alpha^{\dagger}_k,\beta^{\dagger}_k,\alpha_{-k},\beta_{-k}\right)$. Obviously, $H_{\mathrm{BdG}}$ ceases to be Hermitian as a result of the non-Hermitian anti-diagonal terms. The matrix~\eqref{eq:BdG_Hamiltonian} can be diagonalized based on the similarity transformation, 
\begin{eqnarray}
   V^k H_{\mathrm{BdG}} (V^k)^{-1}= \mathrm{diag}(E^k_{++},E^k_{+-},E^k_{--},E^k_{-+}).
\end{eqnarray}
Thus, the corresponding Bogoliubov transformation, which is utilized to diagonalize the non-Hermitian Hamiltonian $H_{k} = E^k_{\rho\sigma} \overline{A}^k_{\rho\sigma} A^k_{\rho\sigma}$, is defined as $A^k_{\rho\sigma} = (V^k\psi_k)_{\rho\sigma}$ and $\overline{A}^k_{\rho\sigma} = [\psi_k(V^k)^{-1}]_{\rho\sigma}$, with $\rho,\sigma = \pm  $.
The specific form of the quasiparticle eigenvalue is given by
\begin{equation}
	E^k_{\rho\sigma} = \rho\sqrt{\Lambda_k+ \sigma \sqrt{\Omega_k}} ,\label{eq:quasiparticle_energy}
\end{equation}
where
\begin{eqnarray}
    \Lambda_k &=& 4\mu ^{2} +4\cos^{2}\frac{k}{2} J^2+4\sin^{2}\frac{k}{2} \Delta ^{2} -4\cos^{2}\frac{k}{2} \delta ^{2} ,  \\
    \Omega_k &=& 64\cos^{2}\frac{k}{2} \mu ^{2} J^{2} -16\sin^{2} k\Delta ^{2} \delta ^{2} -64\cos^{4}\frac{k}{2} \delta ^{2} J^{2} .\notag
\end{eqnarray}
Similar to the operators $(d_j,\overline{d}_j)$ introduced in the previous section, the complex Bogoliubov modes $(\overline{A}^k_{\rho\sigma}, A^k_{\rho\sigma})$ are not related by Hermitian conjugation, i.e., $\overline{A}^k_{\rho\sigma} \ne (A^k_{\rho\sigma})^{\dagger}$, reflecting the non-Hermitian nature of the Hamiltonian. Nevertheless, they satisfy the canonical anti-commutation relations as well,
\begin{eqnarray}
\{A_{\rho \sigma }^{k},\overline{A}_{\rho ^{\prime }\sigma ^{\prime
}}^{k^{\prime }}\} &=&\delta _{kk^{\prime }}\delta _{\rho \rho ^{\prime
}}\delta _{\sigma \sigma ^{\prime }}, \notag \\
\{A_{\rho \sigma }^{k},A_{\rho ^{\prime }\sigma ^{\prime }}^{k^{\prime }}\}
&=&\{\overline{A}_{\rho \sigma }^{k},\overline{A}_{\rho ^{\prime }\sigma
^{\prime }}^{k^{\prime }}\}=0.
\end{eqnarray}
We find that the algebraic nature of the spectrum is governed by $\Omega_k$.
Therefore, the $\mathcal{T}$-symmetry breaking boundary can be determined by the closing of the gap between $E_{\rho+}$ and $E_{\rho-}$ ($\Omega_k=0$), which occurs along the lines $\delta^2 = \mu^2$. The spectrum is real when $\delta < \mu$, whereas complex eigenvalues appear as conjugate pairs once $\delta$ crosses this boundary. 
We plot the $\mathcal{T}$-symmetry-breaking boundary as white solid lines and indicate the TSB region in the two-dimensional plot of the steady order parameter (defined in Eq.~\eqref{eq:order_parameter_staggered}) in Fig.~\ref{fig4}(a).

On the other hand, many-body eigenenergy can be constructed additively from the single quasiparticle spectrum and the ground state is defined as $\vert \mathrm{G}\rangle = \prod_k\overline{A}^k_{–-}\overline{A}^k_{-+}\vert\mathrm{Vac} \rangle$, with the eigenenergy $E_\mathrm{g} = \sum_{\sigma} \sum_{k} E_{-\sigma}^k$. 
The associated biorthogonal state is $\langle\overline{\mathrm{G}}\vert = \overline{\langle \mathrm{Vac} \vert} \prod_k{A}^k_{–-}{A}^k_{-+} $.
Here, $\vert\mathrm{Vac} \rangle$ ($\langle \overline{ \mathrm{Vac} }\vert$) denotes the vacuum state annihilated by the set of operators ${A_{\rho\sigma}^k}$ ($\overline{A}_{\rho\sigma}^k$), i.e., $A_{\rho\sigma} ^k\vert\mathrm{Vac} \rangle = 0$ (${\langle \overline{\mathrm{Vac} }\vert}\overline{A}_{\rho\sigma}^k = 0$).
We find that, irrespective of the algebraic nature of the energy levels, the eigenenergy $E_\mathrm{g}$ remains strictly real. The topological phase boundary can be identified by the closure of the gap between the ground and first excited states, occurring when the second branch $E_{+-}$ and the third branch $E_{--}$ intersect. This condition yields the phase boundary $\delta^2 + J^2 = \mu^2$, which is shown in Fig.~\ref{fig4}(a) by dashed cyan curves.
We introduce the generalized Zak phase $\mathcal{Z}$ in the non-Hermitian case to characterize the topological properties of the ground state,
\begin{eqnarray}
    \mathcal{Z} &=& i\int_0^{\pi}  {\langle\overline{\mathrm{Vac}}\vert}{A}^k_{--} \frac{\partial}{\partial k}\overline{A}^k _{–-}\vert \mathrm{Vac} \rangle  \mathrm{d}k \notag \\
   & + & i\int_0^{\pi}
    {\langle\overline{\mathrm{Vac}}\vert}{A}^k_{–+} \frac{\partial}{\partial k}\overline{A}^k_{-+}\vert \mathrm{Vac} \rangle  \mathrm{d}k.
\end{eqnarray}
We find that when $\delta^2 + J^2 >\mu^2$, the system resides in a topological phase characterized by a nonzero Zak phase, $\mathcal{Z}=\pi$~\cite{zhang2025majorana}, as labeled in Fig.~\ref{fig4}(a); otherwise, $\mathcal{Z}=0$. Unlike the uniform case, the $\mathcal{T}$-symmetry breaking does not affect the ground state or the topological invariant. Meanwhile, a slight staggered imbalance can not only induce imaginary eigenenergies when the onsite potential is small but also influence the phase boundary.

To facilitate analytical and numerical treatment of the time evolution, we represent the momentum-space Hamiltonian $H_k$ within the fermionic Fock space and specify the Hilbert subspace where the quench dynamics take place. 
Due to the parity symmetry $\left[ \Pi ,H\right] =0$, the Hilbert space can be naturally decomposed into even- and odd-parity subspaces. 
Translational symmetry ensures conservation of the total momentum, so that the time evolution is confined to the even-parity subspace with zero total momentum.
We thus decompose the momentum Hamiltonian as $H_k =H_{k, \parallel } \oplus H_{k,\perp}$, where $H_{k, \parallel }$ acts on the subspace relevant to the dynamics of interest, and $H_{k,\perp}$acts on its orthogonal complement. Specifically, the sub-Hamiltonian $ H_{k, \parallel }$ takes the form
\begin{eqnarray}
    H_{k, \parallel } = \varphi^{\dagger}_k h_k\varphi_k,
\end{eqnarray}
\begin{widetext}
where the matrix $h_k$ is given by
    \begin{eqnarray}
    h_k=\left( 
    \begin{array}{ c c c c c c }
0 & \eta _{k} & 0 & 0 & -\eta _{k}^{\ast } & 0\\
-e^{-ik} \eta _{k} & -4\mu  & \gamma _{-k} & \gamma _{-k} & 0 & -\eta _{k}^{\ast }\\
0 & \gamma _{k} & -4\mu  & 0 & \gamma _{-k} & 0\\
0 & \gamma _{k} & 0 & -4\mu  & \gamma _{-k} & 0\\
e^{ik} \eta _{k}^{\ast } & 0 & \gamma _{k} & \gamma _{k} & -4\mu  & \eta _{k}\\
0 & e^{ik} \eta _{k}^{\ast } & 0 & 0 & -e^{-ik} \eta _{k} & -8\mu 
\end{array}
\right) , \label{eq:matrix}
\end{eqnarray}
with the vector basis 
\begin{eqnarray}
     \varphi^{\dagger}=\left(\vert0\rangle_k ,\alpha _{k}^{\dagger } \beta _{-k}^{\dagger } \vert0\rangle_k,\alpha _{k}^{\dagger } \alpha _{-k}^{\dagger } \vert0\rangle_k,\beta _{k}^{\dagger } \beta _{-k}^{\dagger } \vert0\rangle_k ,\beta _{k}^{\dagger } \alpha _{-k}^{\dagger } \vert0\rangle_k ,\alpha _{k}^{\dagger } \beta _{k}^{\dagger } \alpha _{-k}^{\dagger } \beta _{-k}^{\dagger } \vert0\rangle_k \right). \label{eq:dynamical_matrix}
\end{eqnarray}
\end{widetext}
In this case, the normalized evolved state can be written as
\begin{eqnarray}
        \left\vert
\psi \left( t\right) \right\rangle = \frac{\prod_{\pi>k>0} \exp(-iH_{k,\parallel}t)\vert 0  \rangle_k}{ \lVert\prod_{\pi>k>0} \exp(-iH_{k,\parallel}t)\vert 0  \rangle_k \rVert}.
\end{eqnarray}
$\lVert \cdot\rVert$ denotes the norm of the state. 
Based on the structure of the vector basis, we define a specific operator, distinct from the form in Eq.~\eqref{eq:orderparameter}, that captures the pairing rate in momentum space for a system with two sublattices, i.e.,
\begin{equation}
\widehat{\mathcal{O}}_{k}^{\gamma\gamma'}= \sum_{\gamma,\gamma' \in \{\alpha,\beta\}}i \gamma _{k}^{\dagger } {\gamma'}_{-k}^{\dagger } +\mathrm{h.c.}.
\end{equation}
It characterizes intra-sublattice pairing when $\gamma = \gamma’$, and inter-sublattice pairing when $\gamma \neq \gamma’$. The dynamical order parameter and its steady value are given by
\begin{eqnarray}
    \mathcal{O}(t) &=& \frac{1}{N} \sum_{k,\gamma,\gamma'}\vert \langle \psi(t)\vert \widehat{\mathcal{O}}_{k}^{\gamma\gamma'}\vert \psi(t)\rangle \vert, \notag \\
    \overline{\mathcal{O}} &=& \lim_{T \rightarrow\infty}\frac{1}{T}\int_0^T \mathcal{O}(t)\mathrm{d}t .\label{eq:order_parameter_staggered}
\end{eqnarray}
Fig.~\ref{fig4} shows numerical results of the steady order parameter $\overline{\mathcal{O}}$ in the $\mu$–$\delta$ plane. 
As expected, the quantity $\overline{O}$ exhibits non-analytic behavior along both the $\mathcal{T}$-symmetry-breaking boundary and the topological phase boundary, closely resembling the Hermitian Kitaev chain discussed in Sec.~\ref{Hermitian Kitaev chain and pairing in $k$ space} and the uniform case in Sec.~\ref{Uniform pairing imbalance}. 
However, within the TSB regions, the order parameter increases sharply as $|\mu|$ decreases, reaching a stable maximum at zero chemical potential, in contrast to the plateau observed in the other two cases.
These phenomena can be explained as follows. For a non-Hermitian Hamiltonian with complex eigenvalues, the long-time evolved state is dominated by the eigenstate corresponding to the largest imaginary part of the eigenenergy, reflecting the critical dynamics. 
As a result, within the topological and TSB regions, the order parameter accurately captures the characteristics of this dominant eigenstate. 
Remarkably, at the symmetry point $\mu = 0$ (excluding the Hermitian case $\delta = 0$), the steady order parameter attains its maximum value, which remains essentially constant as $\delta $ varies [cf. Fig.~\ref{fig4}(b)].

\begin{figure}[t!]
\centering\includegraphics[width=0.4\textwidth]{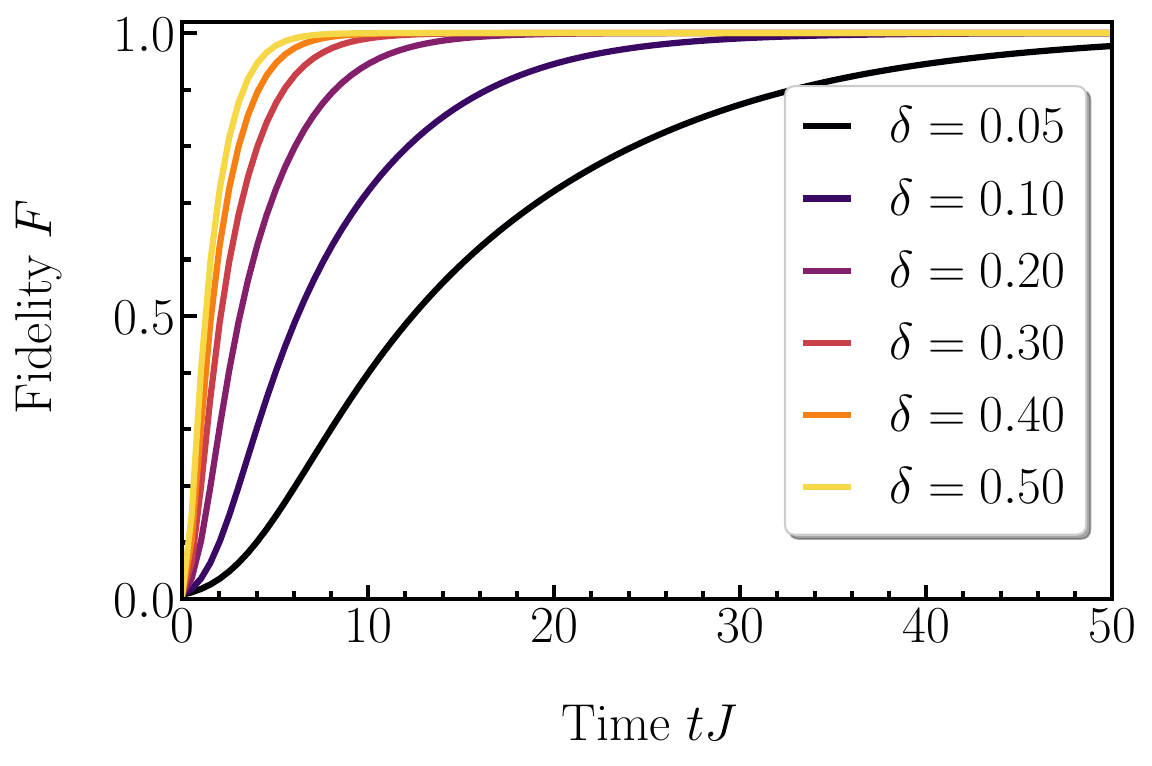}
\caption{Profiles of the fidelity $F$, obtained from simulation of quench dynamics in real space, as a function of time $t$ for different values of $\delta$. The results show that, regardless of the amplitude of the staggered imbalance, the system eventually evolves to the same eigenstate, whose eigenvalue has the largest imaginary part. Other parameters are $J=1$, $N= 10$, $\mu=0$, $\Delta = 1$.}
\label{fig5}
\end{figure}

\begin{figure*}[t!]
\centering\includegraphics[width=0.9\textwidth]{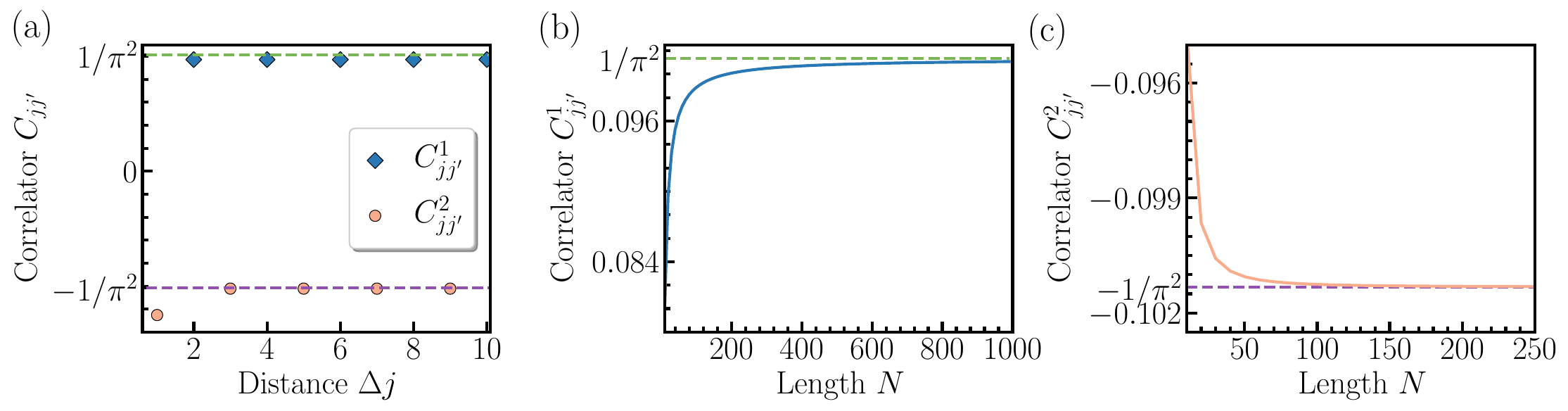}
\caption{(a) Numerical results of the correlator, obtained via exact diagonalization, are shown as a function of the distance $\Delta j = \vert j - j'\vert \in [0,N]$ for a system of size $N=10$. The results indicate that $C_{jj'}^1$ (blue diamond) and $C_{jj}^2$ (orange cycle) converge to $\frac{1}{\pi^2}$ (green dashed line) and $-\frac{1}{\pi^2}$ (purple dashed line), respectively, as the distance increases. 
The correlator is plotted as a function of system size $N$ for 
(b) odd distances $C_{jj’}^{1}$ (blue curve, Eq.~\eqref{Cjj1}) and (c) even distances $C_{jj’}^{2}$ (orange curve, Eq.~\eqref{Cjj2}). The difference between the numerical results and predicted values decreases gradually as the system size increases. We set the distance as the half of system size.}
\label{fig6}
\end{figure*}

\section{Dynamical generation of superconducting state with ODLRO}
\label{Dynamical generation of superconducting state with ODLRO}

In this section, we study the properties of the eigenstates at symmetry point $\mu = 0$. 
We find that the system hosts two eigenstates whose configurations are independent of the specific system parameters. 
These eigenstates possess conjugate imaginary energies, with magnitudes proportional to the imbalanced strength $\delta$. 
Furthermore, both eigenstates can be expressed as the condensation of pairs and exhibit possess nonzero long-range pair-pair correlation.

Referring to Eq.~\eqref{eq:quasiparticle_energy}, when $\mu = 0$, the quasi-particle eigenvalues are in the form of
\begin{eqnarray}
    E^k_{\rho\sigma} = 2\rho\sqrt{J^2\cos^2\frac{k}{2}+\Delta^2 \cos^2\frac{k}{2}} + 2i \sigma \rho \delta \cos\frac{k}{2}. 
\end{eqnarray}
Thus, when $\delta$ is positive, the many-body eigenvalue with the largest imaginary part is
$
    \varepsilon_{+} = 4i\delta\cos\frac{k}{2},
$
and the corresponding eigenstate takes the form
\begin{eqnarray}
    \vert \Phi_+ \rangle = \prod_k \overline{A}_{++}^k \overline{A}_{--}^k \vert \mathrm{Vac}\rangle.
\end{eqnarray}
This eigenstate can also be expressed in the Fock basis as
\begin{eqnarray}
\vert  \Phi_+ \rangle  =
\prod_k \Gamma_{k,+}\Gamma_{-k,+} \vert0\rangle_k,
\end{eqnarray}
where
\begin{eqnarray}
\Gamma_{k,\sigma} =\frac{1}{\sqrt{2}} \left[ 1 -  \sigma i \exp\left(-i\frac{k}{2}\right) \alpha _{k}^{\dagger}\beta _{-k}^{\dagger} \right].
\end{eqnarray}
Correspondingly, when $\delta$ is negative, the eigenvalue with the largest imaginary part becomes $\varepsilon_{-} = -4i\delta\cos\frac{k}{2}$, and the corresponding eigenstate is given by
\begin{eqnarray}
    \vert  \Phi_- \rangle &= &\prod_k  \Gamma_{k,-}\Gamma_{-k,-}\vert0\rangle_k,
\end{eqnarray}
serving as the counterpart of $\varepsilon_+$ for negative $\delta$. The results can be obtained by diagonalized the matrix~\eqref{eq:dynamical_matrix} as well.
Therefore, the characteristics of the evolution become clear:
When $\delta > 0$, the system evolves towards a final state dominated by the exponential growth of the coefficients $\exp(\varepsilon_{+} t)$. In this case, the asymptotic state is
$
\vert\psi_k(t \rightarrow \infty)\rangle = \vert \Phi_+ \rangle.
$
Fig.~\ref{fig5} presents numerical simulations of the fidelity, defined as the squared modulus of the overlap between the evolved state and the target state, i.e.,
\begin{eqnarray}
F(\delta,t) =\left\vert\frac{\langle \Phi_+\vert \exp[-iH(\mu=0,\delta)t]\vert 0\rangle   }{\lVert \exp[-iH(\mu=0,\delta)t]\vert 0\rangle\rVert }  \right\vert^2.
\end{eqnarray}
We find that, irrespective of the imbalanced strength, the fidelity approaches unity at long times, consistent with our analytical results.
Conversely, when $\delta < 0$, the system evolves towards the alternative asymptotic state
$
\vert \psi_k(t \rightarrow \infty) \rangle = \vert \Phi_- \rangle.
$
The dynamics become highly sensitive to the sign of the imbalanced strength due to the presence of the imaginary eigenvalues at the symmetry point.

In the following, we show that these two eigenstates can be expressed as superpositions of fermion pair-condensate states, analogous to the $\eta$-pairing states in the Hubbard model~\cite{Yang1989eta}, with varying numbers of pairs. This representation suggests that these states correspond to superconducting phases.
As is well known, fermions obey the Pauli exclusion principle, prohibiting them from occupying the same quantum state, unlike bosons which can all condense into the same ground state. However, fermion pairs can, to some extent, behave as bosons. To characterize their properties, we introduce the pairing correlation function $\hat{C}_{jj'}$. We find that both states exhibit ODLRO. Taking $\vert \Phi_-\rangle$as an example, the results apply equally to $\vert \Phi_+\rangle$. The collective BCS-pair operators are then defined as
\begin{eqnarray}
	s^{+} &=&\left( s^{-}\right) ^{\dagger } \\
	&=&i\sum_{k\in (0,\pi )}\left(e^{-i\frac{k}{2}}\alpha _{k}^{\dagger }\beta
	_{-k}^{\dagger }+e^{i\frac{k}{2}}\alpha _{-k}^{\dagger }\beta _{k}^{\dagger }\right) .\notag
\end{eqnarray}
Here, $s^{+}$ acts as a collective pair-creation operator which, when applied to the vacuum state $s^{+}\vert0\rangle$, generates fermion pairs each consisting of one fermion on the odd sublattice and one on the even sublattice, such that each pair carries zero total momentum.
These pairs resemble BCS-like pairings formed across the two sublattices. Accordingly, the $s^z$ takes the form 
\begin{equation}
	s^{z}=\frac{1}{2}\sum_{k\in (-\pi ,\pi )}(\alpha _{k}^{\dagger }\alpha
	_{k}+\beta _{k}^{\dagger }\beta _{k}-1).
\end{equation}
These pseudo spin operators satisfying the Lie algebra commutation relations, i.e.,
\begin{equation}
	\left[ s^{+},s^{-}\right] =2s^{z},\left[ s^{z},s^{\pm }\right] =\pm s^{\pm }.
\end{equation}
We find that $\vert \Phi_-\rangle$ can be represented as a superposition of pair creation condensate state, i.e.,
\begin{equation}
	\vert \Phi_-\rangle = \frac{1}{\sqrt{N}}
	\sum_{n=0}^{N}\frac{1}{\Omega_n}\left( s^{+}\right) ^{n}\left\vert
	0\right\rangle ,
\end{equation}
with $\Omega_n = n! \sqrt{C_N^n}$, which is a signal of superconductivity. To prove this, we introduce the correlation operator $\hat{C}_{jj^{\prime }} =  c^{\dagger }_{j^{\prime}}c^{\dagger }_{j^{\prime }+1}  c_{j+1}c_{j}$. $\langle\hat{C}_{jj^{\prime }} \rangle$ measures the correlation of two pairs with distance $\Delta j = \vert j-j'\vert$.

To simplify the discussion, we consider $j'$ to be odd. The correlator $\langle\hat{C}_{jj^{\prime }} \rangle$ then admits two analytic forms, depending on the parity of $\Delta j$, i.e.,
\begin{eqnarray}
    \langle\hat{C}_{jj^{\prime }} \rangle = 
    \begin{cases}
C_{jj'}^1, & j'=2l'-1\textrm{ and }j=2l-1, \\
C_{jj'}^2,  & j'=2l'-1\textrm{ and }j=2l.
\end{cases}
\end{eqnarray}
Applying the operator $c_{2l}c_{2l-1}$ to the state $\vert \Phi_-\rangle$, we obtain
\begin{widetext}
\begin{eqnarray}
c _{2l}c_{2l-1}\left\vert \Phi_-\right\rangle &=&
\frac{1%
}{2^{N}N} \bigg\{
\sum_{k_{1},k_{2}}
e^{i(k_1+k_2)l} 
e^{\frac{i}{2}(k_2-k_1)}
\alpha _{-k_{2}}^{\dagger }\beta
_{-k_{1}}^{\dagger } 
 \prod_{\pi >k>-\pi }^{k\neq k_{1},-k_{2}}\left(1+ie^ {-\frac{ik}{2}} \alpha _{k}^{\dagger }\beta _{-k}^{\dagger }\right)\left\vert 0\right\rangle  
\notag \\
&+&\sum_{k_{1}}ie^ {-\frac{ik_1}{2}}
\prod_{\pi >k>-\pi }^{k\neq
k_{1}}\left[1+ie^{-\frac{ik}{2}}\alpha _{k}^{\dagger }\beta _{-k}^{\dagger }\right]\left\vert
0\right\rangle \bigg\} .
\end{eqnarray}
In a similar manner, when we apply $c_{2j+1}c_{2j}$ to the state, we obtain
\begin{eqnarray}
c_{2j+1}c _{2j}\left\vert \Phi_-\right\rangle &=& -
\frac{%
1}{2^{N}N}\bigg\{\sum_{k_{1},k_{2}}
e^{i(k_1+k_2)j}
e^{\frac{i}{2}(k_2+k_1)}
\alpha _{-k_{2}}^{\dagger }\beta
_{-k_{1}}^{\dagger } 
\prod_{\pi >k>-\pi }^{k\neq k_{1},k_{2}}\left(1+i
e^{-\frac{ik}{2}}
\alpha _{k}^{\dagger }\beta _{-k}^{\dagger }\right)\left\vert 0\right\rangle 
\notag \\
&&+\sum_{k_{1}}i
e^{\frac{ik_1}{2}}
\prod_{\pi >k>-\pi }^{k\neq
k_{1}}
\left(1+i
e^{-\frac{ik}{2}}
\alpha _{k}^{\dagger }\beta _{-k}^{\dagger }\right)
\left\vert
0\right\rangle \bigg\}  .
\end{eqnarray}
\end{widetext}
A straightforward derivation yields,
\begin{eqnarray}
C_{jj^{\prime }}^{1} &=& \frac{1}{4N^{2}}\sum_{k_{1},k_{2}\neq
-k_{1}}e^{i\left( k_{1}+k_{2}\right) \left( l-l^{\prime }\right) } \notag \\
&+&\frac{1}{%
4N^{2}}\sum_{k_{1},k_{2}}e^{i\frac{k_{1}-k_{2}}{2}},  \label{Cjj1} \\
C_{jj^{\prime }}^{2}&=&-\frac{1}{4N^{2}}\sum_{k_{1},k_{2}\neq
-k_{1}}e^{i\left( k_{1}+k_{2}\right) \left( l-l^{\prime }\right) -ik_{1}} \notag \\
&-&%
\frac{1}{4N^{2}}\sum_{k_{1},k_{2}}e^{-i\frac{k_{1}+k_{2}}{2}}. \label{Cjj2} 
\end{eqnarray}
In the limit $N\rightarrow\infty$ and for $\lvert j-j’\rvert \gg 1$, the rapidly oscillating terms vanish, yielding
\begin{eqnarray}
{C}_{jj^{\prime }}^{1} =\frac{1}{\pi ^{2}} ,  \quad {C}_{jj^{\prime }}^{2}  =-\frac{1}{\pi ^{2}}.
\end{eqnarray}
Fig.~\ref{fig6}(a) presents the correlation as a function of the distance $\Delta j=\lvert j-j^{\prime} \rvert$ for a system of size $N = 10$, in agreement with our analytical prediction. Figs.~\ref{fig6}(b) and \ref{fig6}(c) display the dependence of $C_{jj’}^1$ and $C_{jj’}^2$ on the system size, respectively, highlighting the finite-size effects.
We therefore conclude that critical dynamics can generate two distinct superconducting states exhibiting exact ODLRO at the symmetry point, even for infinitesimal perturbations, where the realized state is dictated by the direction of the bias.
The proposed dynamical scheme can be realized in systems that are well within the capabilities of current experimental techniques~\cite{xu2020probing,shen2025observation,Fleckenstein2022non}. 
Consequently, it provides a distinctive route, rather than cooling down the temperature~\cite{marti2025efficient} or imaginary time evolution~\cite{Real2021Lin}, to generating a steady superconducting state far from the ground state.
In contrast, for other regions of the phase diagram, no definitive conclusion can be drawn due to the absence of exact analytical results. 
\section{Summary}
\label{Summary}
In summary, we investigate the equilibrium and nonequilibrium properties of a Kitaev chain with non-Hermitian pairing imbalance. We first introduce the properties of Hermitian Kitaev chain as a baseline for comparison. In this case, a certain order parameter of the evolved state becomes steady after a relaxation time. The steady value can be used to determine the phase boundary, which constitutes one notion of a dynamical phase transition. In the topological phase, the order parameter is independent of the onsite potential, exhibiting a plateau structure.
Based on a similarity transformation, we find that a uniform imbalance neither changes the phase boundary nor the algebraic nature of the spectrum, thereby preserving most of the conclusions under this condition. The results show that although the non-Hermitian term is introduced, the evolved state undergoes a coherent dynamics, leading to an intrinsic balance.

However, a slight staggered imbalance not only modifies the topological phase boundary but also breaks the $\mathcal{T}$ symmetry when the onsite potential is small. Although the steady order parameter still signals the phase boundary, its behavior changes qualitatively. In the TSB phase, the system evolves into a single eigenstate associated with the largest imaginary eigenenergy. At the symmetry point, we analytically compute this eigenstate and demonstrate that it exhibits exact ODLRO based on the pair-pair correlator. 
Our results emphasize the fragility of critical nonequilibrium dynamics in non-Hermitian topological systems and elucidate the interplay among non-Hermiticity, topology, and dynamical criticality.
Moreover, we also suggest a simple and efficient mechanism to prepare superconducting states far from the ground state. 
In future work, we plan to extend our investigation to OBC, where translational symmetry is broken but Majorana edge modes appear at the system boundaries. 
A potential research topic is to explore whether several exotic states in OBC, such as Majorana edge modes in minimal Kitaev chains at sweet spots~\cite{Dvir2023Realization, cayao2025non} and nonendpoint Majorana bound states in extended Kitaev chains~\cite{zhang2024nonendpoint}, can be prepared via similar schemes.

\section{Acknowledgment}
Y. B. Shi and Y. X. Zhang thank S. Zhang, X. Z. Zhang, Y. M. Hu, R. Qi, and R. Y. Yin for enlightening discussions. Y. B. Shi and S. W. Liu appreciate the support from H. Z. Zhao. We acknowledge the support of NSFC (Grants No. 12374461).

\bibliography{reference}

\end{document}